\documentclass[12pt,preprint]{aastex}
\begin{document}
\title{
A Northern Sky Survey for Steady TeV Gamma-Ray Point Sources Using
the Tibet Air Shower Array}
\author{ M.~Amenomori\altaffilmark{1},
  S.~Ayabe\altaffilmark{2},
  D.~Chen\altaffilmark{3},
  S.W.~Cui\altaffilmark{20},
  Danzengluobu\altaffilmark{5},
  L.K.~Ding\altaffilmark{4},
  X.H.~Ding\altaffilmark{5},
  C.F.~Feng\altaffilmark{6},
  Z.Y.~Feng\altaffilmark{7},
  X.Y.~Gao\altaffilmark{8},
  Q.X.~Geng\altaffilmark{8},
  H.W.~Guo\altaffilmark{5},
  H.H.~He\altaffilmark{4},
  M.~He\altaffilmark{6},
  K.~Hibino\altaffilmark{9},
  N.~Hotta\altaffilmark{10},
  Haibing Hu\altaffilmark{5},
  H.B.~Hu\altaffilmark{4},
  J.~Huang\altaffilmark{11},
  Q.~Huang\altaffilmark{7},
  H.Y.~Jia\altaffilmark{7},
  F.~Kajino\altaffilmark{12},
  K.~Kasahara\altaffilmark{13},
  Y.~Katayose\altaffilmark{3},
  C.~Kato\altaffilmark{14},
  K.~Kawata\altaffilmark{11},
  Labaciren\altaffilmark{5},
  G.M.~Le\altaffilmark{15},
  J.Y.~Li\altaffilmark{6},
  H.~Lu\altaffilmark{4},
  S.L.~Lu\altaffilmark{4},
  X.R.~Meng\altaffilmark{5},
  K.~Mizutani\altaffilmark{2},
  J.~Mu\altaffilmark{8},
  K.~Munakata\altaffilmark{14},
  A.~Nagai\altaffilmark{16},
  H.~Nanjo\altaffilmark{1},
  M.~Nishizawa\altaffilmark{17},
  M.~Ohnishi\altaffilmark{11},
  I.~Ohta\altaffilmark{10},
  H.~Onuma\altaffilmark{2},
  T.~Ouchi\altaffilmark{9},
  S.~Ozawa\altaffilmark{11},
  J.R.~Ren\altaffilmark{4},
  T.~Saito\altaffilmark{18},
  M.~Sakata\altaffilmark{12},
  T.~Sasaki\altaffilmark{9},
  M.~Shibata\altaffilmark{3},
  A.~Shiomi\altaffilmark{11},
  T.~Shirai\altaffilmark{9},
  H.~Sugimoto\altaffilmark{19},
  M.~Takashima\altaffilmark{12}
  M.~Takita\altaffilmark{11},
  Y.H.~Tan\altaffilmark{4},
  N.~Tateyama\altaffilmark{9},
  S.~Torii\altaffilmark{9},
  H.~Tsuchiya\altaffilmark{11},
  S.~Udo\altaffilmark{11},
  T.~Utsugi\altaffilmark{9},
  H.~Wang\altaffilmark{4},
  X.~Wang\altaffilmark{2},
  Y.G.~Wang\altaffilmark{6},
  H.R.~Wu\altaffilmark{4},
  L.~Xue\altaffilmark{6},
  Y.~Yamamoto\altaffilmark{12},
  C.T.~Yan\altaffilmark{4},
  X.C.~Yang\altaffilmark{8},
  S.Yasue\altaffilmark{14},
  Z.H.~Ye\altaffilmark{15},
  G.C.~Yu\altaffilmark{7},
  A.F.~Yuan\altaffilmark{5},
  T.~Yuda\altaffilmark{11},
  H.M.~Zhang\altaffilmark{4},
  J.L.~Zhang\altaffilmark{4},
  N.J.~Zhang\altaffilmark{6},
  X.Y.~Zhang\altaffilmark{6},
  Yi~Zhang\altaffilmark{4},
  Y.~Zhang\altaffilmark{4},
  Zhaxisangzhu\altaffilmark{5},
  and X.X.~Zhou\altaffilmark{7}\\
(The Tibet AS${\bf \gamma}$ Collaboration) }

\altaffiltext{1}{\footnotesize Department of Physics, Hirosaki University, Hirosaki 036-8561, Japan}
\altaffiltext{2}{Department of Physics, Saitama University, Saitama 338-8570, Japan}
\altaffiltext{3}{Faculty of Engineering, Yokohama National University, Yokohama 240-8501, Japan}
\altaffiltext{4}{Key Laboratory of Particle Astrophysics, Institute of High Energy Physics, Chinese Academy of
Sciences, Beijing 100049, China}
\altaffiltext{5}{Department of Mathematics and Physics, Tibet University, Lhasa 850000, China}
\altaffiltext{6}{Department of Physics, Shandong University, Jinan 250100, China}
\altaffiltext{7}{Institute of Modern Physics, South West Jiaotong University, Chengdu 610031, China}
\altaffiltext{8}{Department of Physics, Yunnan University, Kunming 650091, China}
\altaffiltext{9}{Faculty of Engineering, Kanagawa University, Yokohama 221-8686, Japan}
\altaffiltext{10}{Faculty of Education, Utsunomiya University, Utsunomiya 321-8505, Japan}
\altaffiltext{11}{Institute for Cosmic Ray Research, University of Tokyo, Kashiwa 277-8582, Japan}
\altaffiltext{12}{Department of Physics, Konan University, Kobe 658-8501, Japan}
\altaffiltext{13}{Faculty of Systems Engineering, Shibaura Institute of Technology, Saitama 330-8570, Japan}
\altaffiltext{14}{Department of Physics, Shinshu University, Matsumoto 390-8621, Japan}
\altaffiltext{15}{Center of Space Science and Application Research, Chinese Academy of Sciences, Beijing 100080,
China}
\altaffiltext{16}{Advanced Media Network Center, Utsunomiya University, Utsunomiya 321-8585, Japan }
\altaffiltext{17}{National Institute for Informatics, Tokyo 101-8430, Japan}
\altaffiltext{18}{Tokyo Metropolitan College of Aeronautical Engineering, Tokyo 116-0003, Japan}
\altaffiltext{19}{Shonan Institute of Technology, Fujisawa 251-8511, Japan}
\altaffiltext{20}{Department of Physics, HeBei Normal University, Shijiazhuang 050016, China}

\begin{abstract} 
Results on steady TeV $\gamma$-ray point source search using data taken from the Tibet HD (Feb. 1997 -- Sep. 1999) and Tibet III (Nov. 1999 -- Oct. 2001) arrays are presented. From $0^{\circ}$ to $60^{\circ}$ in declination, significant excesses from the well-known steady source Crab Nebula and the high state of the flare type source Markarian 421 are observed. Because the levels of significance from other positions are not sufficiently high, 90\% confidence level upper limits on the flux are set assuming different power law spectra. To allow crosschecking, two independently developed analyses are used in this work.
\end{abstract}


\section{Introduction}

In the last 15 years, Imaging Atmospheric Cherenkov Telescopes (IACT) have discovered more than 10
sources in TeV energy range populating from pulsar nebula, supernova remnants, starburst
galaxies, AGNs of BL Lac type and even a newly found unidentified source \cite{Ong2003, Horan2004, Weeks2003, Voelk2003}, 
and it is remarkable that AGNs make up a large percentage of these sources.
To understand the mechanism of the TeV $\gamma$-ray emissions from diversity of
objects, and more fundamentally, to understand the origin and acceleration
of cosmic ray (CR), it is important that the number of TeV sources will be largely
increased.  However,
 to improve the statistics of TeV sources is not a trivial task for IACT due to its small field of view
and low duty cycle, which particularly limit its power in detecting
 flare type sources and unknown sources. Therefore, a complementary detecting technology is necessary. 
Although their sensitivity for detecting dim $\gamma$-ray sources is lower than that of the IACT, 
Extensive Air Shower (EAS) experiments, such as the Tibet air shower array \cite{Amenomori1999, Amenomori2000, Amenomori2003} and 
Milagrito/Milagro 
(Atkins et al. 1999;  Atkins et al. 2003; Atkins et al. 2004) experiments, have resulted in successful observation of $\gamma$-ray emissions from 
standard candle Crab nebula and from transient sources such as Mrk~501 and Mrk~421. Their characteristic abilities 
in high duty cycle and large field of view allow them to simultaneously monitor a larger area 
in space over continuous time. Most importantly, the EAS technique can potentially provide
useful
information for 
IACT observatories for their further dedicated observations. Such follow-up observations have been performed by 
Whipple experiment \cite{Falcone2003a, Walker2003b} guided by the hot spots seen from TeV $\gamma$-ray all-sky 
survey results published by the Tibet air shower array \cite{Amenomori2001a, Cui2003} and Milagro \cite{Sinnis2002} experiments.

The satellite experiment EGRET on board the Compton Gamma Ray Observatory (CGRO) pioneered the all-sky $\gamma$-ray 
survey in the energy range of 20 MeV to 30 GeV \cite{Hartman1999}. The successful detection of the diffuse 
$\gamma$-ray emissions from the galactic plane and a peak near 70 MeV seen in a rather hard $\gamma$-ray spectrum
supported the supernova acceleration model, according to which TeV $\gamma$-ray
emissions were also expected. In addition, EGRET discovered 271 $\gamma$-ray point sources \cite{Hunter1997}, among
which about 100 were identified as SNRs, AGNs, etc., while the rest has remained unidentified. The diverse types of  
$\gamma$-ray emissions, which included galactic sources and extragalactic
sources, greatly improved our knowledge of both CR physics and astrophysics.

The all-sky survey remains one of the major concerns for ground-based observatories (Weeks et al. 1989; Alexandreas et al. 1991; Mckay et al. 1993; Wang et al. 2001; Aharonian et al. 2002a; Aharonian et al. 2002b; Atkins et al. 2004; Antoni et al. 2004).
For example, AIROBICC  made a northern sky survey \cite{Aharonian2002a} on $\gamma$-ray emissions for energy above 15 TeV; because there was no compelling evidence for a signal source, an absolute flux upper limit between 4.2 and 8.8 Crab 
was obtained for a declination (Decl.) of $0^{\circ}$ and $60^{\circ}$. With the IACT technique, within the 
TeV energy range, the HEGRA telescope was used to performe a survey
 \cite{Aharonian2002b} in one quarter of the galactic plane in the longitude range 
from $-2^{\circ}$ to $85^{\circ}$. As a result of negative evidence, upper flux limits on each individual 
known point source located in this region were obtained between 0.15 Crab and several Crab. Most recently, 
Milagro updated its results \cite{Atkins2004} in the northern sky survey in TeV energy range, and pushed the average flux upper 
limit down to a level between 275 and 600 mcrab ($4.8\times 10^{-12}$ to
 $10.5 \times 10^{-12}{\rm cm}^{-2}{\rm s}^{-1}$) above 1 TeV for source Decl.
between $5^{\circ}$ and $70^{\circ}$.
As an independent experiment, the Tibet air shower array has similar sensitivity, and covers almost the same energy ranges and field of view as those experiments do. The results from the Tibet air shower array should therefore provide important information for crosschecking and confirmation.

\section{Tibet Air Shower Observatory}

The Tibet air shower array has been conducted at Yangbajing ($90.522^{\circ}$E, $30.102^{\circ}$N; 4,300~m a.s.l) in Tibet, China since 1990. The Tibet I array \cite{Amenomori1992}, which consisted of 49 scintillation counters
forming a $7\times7$ matrix of 15~m span, was gradually expanded to the Tibet II array occupying an area of 36,900~m$^2$ by increasing the number of counters from 1994 to 1996. Both Tibet I and II have the same mode energy (the most probable gamma ray energy observed from the standard candle Crab) of 10 TeV. In order to observe TeV CRs, in 1996, part of the Tibet II array covering an area of 5,175~m$^2$ was further upgraded to a high density (HD) array \cite{Amenomori2001b} with 7.5 m span. Soon after decreasing the mode energy to 3 TeV, the HD array observed multi-TeV $\gamma$-ray signals from the Crab Nebula \cite{Amenomori1999} and Mrk~501 \cite{Amenomori2000}. To increase the event rate, in 1999, the HD array was enlarged to cover the center part of the Tibet II array as the Tibet III array \cite{Amenomori2001c}. 
The area of Tibet III has reached to 22,050~m$^{2}$ (Figure 1).

As can be shown in Figure 1, the Tibet HD and Tibet III have identical
structure except the array size and shape.  Each counter has a plastic
scintillation plate (BICRON BC-408A) of 0.5 m$^{2}$ in area and 3 cm in
thickness, and is equipped with a fast-timing (FT) photomultiplier tube
(PMT; Hamamatsu H1161). A 0.5 cm thick lead plate is put on the
top of each counter in order to increase the array sensitivity by converting
$\gamma$-rays into electron-positron pairs in the shower
 \cite{Bloomer1988,Amenomori1990}.
The angular resolution of the HD and the Tibet III are about $0.9^\circ$ in the energy region above 3 TeV, as estimated from full Monte Carlo (MC) simulation \cite{Amenomori1990, Kasahara2003} and verified by the moon shadow measurement from observational data \cite{Amenomori1993,Amenomori2001c,Amenomori2003}.
The trigger rate is about 105 Hz for the HD and 680 Hz for the Tibet III. In this work, the sample includes data obtained by running the HD array for 555.9 live days from Feb. 1997 to Sept. 1999 and the Tibet III array for 456.8 days from Nov. 1999 to Oct. 2001. 

The event selection was done by imposing the following four criteria
on the reconstructed data:
i) each shower event should fire four or more Fast-Time (FT) detectors recording 1.25 or more
   particles;
ii) the estimated shower center location should be inside the array;
iii) the sum of the number of particles per m$^2$ detected in each
detector $\sum \rho_{\rm {FT}}$ should be larger than $15$;
iv) the zenith angle of the incident direction should be less than $40^\circ$.
After applying these cuts and a data quality cut,
about 40\% of the shower events were selected, results in a
total number of events
to be
 about $7.0\times10^9$.

\section{Analysis} \label{ANAL}

Because the Tibet air shower array can not distinguish a $\gamma$-ray induced shower event from the overwhelming CR background shower events, when tracing and counting the number of events in an ``on-source window'' centered at a candidate point source direction with a size at the level of angular resolution, the number of background events must be estimated from the observational data recorded in the side band, which is usually referred to as ``off-source window''.

Sitting on an almost horizontal plane, the Tibet HD and III have
almost azimuth-independent efficiency in receiving
the shower events
for any given
zenith angle. The equi-zenith angle method was therefore developed. In brief, simultaneously collected shower events in the same zenith angle belt can be used to construct the ``off-source windows'' and to estimate the background for a candidate point source located in the same zenith angle. This method can eliminate various detecting effects caused by instrumental and environmental variations, such as changes in pressure and temperature which are hard to be
controlled and intend to introduce systematic error in measurement.
In reality, the actual azimuth angle distribution deviates from a uniform one mainly due to the fact that the Tibet HD and III arrays are situated on a slope of about $1.3^{\circ}$ from the southeast to the northwest. Furthermore, a geomagnetic effect in the northern hemisphere causes unequal efficiency between north and south \cite{Ivanov1999}. Together with other possible unknown effects, all the steady effects lead to an approximately $2.5\%$ higher event rate from the southerly direction than from the northerly direction. This non-uniformity must be properly accounted for in the following analyses.

For the purpose of crosschecking, two independently developed analyses based on the equi-zenith angle method are used in this work. One is dedicated to a point source search, while the other in addition is sensitive to large-scale anisotropy of the intensity of the CR. A MC study was used to carefully compare the two methods, and the search sensitivities were found to be very comparable. However,
the number of the background events estimated by the two methods is different due to the different ways used in choosing the off-source windows (see subsections below in detail).  As a result, significance
values calculated from the two methods differ by about 0.5 $\sigma$ statistically.

\subsection{Method I (short distance equi-zenith angle method)}

As a good approximation, the large-scale anisotropy of the CR intensity can be neglected if the ``off-source windows'' are chosen to be close enough in distance to the ``on-source window'', since the CR intensity only changes slowly in the sky. In this analysis, to make sure not to miss any possible unknown source, the surveyed sky has been over-sampled by the following way: the sky is divided into $0.1^{\circ} \times 0.1^{\circ}$ cells, from $0^{\circ}$ to $360^{\circ}$ in Right Ascension (R.A.) and from $0^{\circ}$ to $60^{\circ}$ in Decl. Pointing to the center of those cells, cones with a half opening angle of $0.9^{\circ}$ (for ${\rm E} > 3$ TeV) or $0.4^{\circ}$ (for ${\rm E} > 10$ TeV) are tested as on-source windows. These cone sizes are chosen to maximize the source search sensitivity \cite{Chen2004}. 10 off-source windows of the same shape are symmetrically aligned on both sides of the on-source window, at a zenith angle-dependent step size in order to maintain an angular distance of $2^{\circ}$ between the neighboring off-source windows. In order to avoid signal events related to an on-source window being wrongly counted as a background event, the two off-source windows closest in distance to the on-source window are required to be placed at twice of the step-size from the center of the on-source window. Denoting the number of events in on-source window as $N_{\rm {ON}}$, and the number of events in the $i$th off-source window as $N_{\rm {OFF} \it i}$, without considering the effect from non-uniform azimuth angle distribution, the number of background events can be estimated as
\begin{equation}
N_{\rm {BG}}=\frac{\sum_{i=1}^{10} N_{\rm {OFF} \it
i}}{10} \label{1111}
\end{equation}
When the zenith angle is less than $4.2^{\circ}$, the 10 off-source windows defined above become to be overlapped. Therefore, data obtained with a zenith angle less than $4.2^{\circ}$ are not used. Since the azimuth angle distribution is not completely flat as discussed earlier, the averaged number of events from off-source windows, $ N_{\rm {BG}}$, as calculated in equation (1) is not an exact description of the background number in the on-source window. Therefore $ N_{\rm {BG}}$ needs to be corrected.
Giving the fact that azimuth angle distribution remains stable over a period
of time that is much longer than the time scale of a day, the effect of the tiny
violation of the exact equi-zenith angle condition can thus be determined
from the observational data which were taken only a few hours apart from
the exposure of the on (off)-source window(s). 
 As a matter of fact,
 the previous measurement is repeated 35 times (the relevant ``on'' and ``off'' source windows are hereafter called ``dummy on'' and ``dummy off'' source windows). Together with the on-source window, 35 dummy on-source windows on the orbit which has the same Decl. as the on-source window are defined at regular intervals along R.A. direction, i.e., the dummy on-source windows are located at $\alpha_{\rm {ON}}+10^{\circ} \times j$, where $\alpha_{\rm {ON}}$ is the R.A. location of the on-source window, and $j$ an index of the repeated measurements which runs from 1 to 35. $N^{'}_{\rm {ON} \it j}$ and $N^{'}_{\rm {BG} \it j}$ represent the number of events in the dummy on-source window and its estimated background from the $j$th measurement, respectively. The correction factor, $\eta$, due to non-uniform azimuth angle distribution is given by
\begin{equation}
{\eta}=\frac{\sum_{j=1}^{35}{N^{'}_{\rm {ON \it j}}}/{35}}
{\sum_{j=1}^{35} {N^{'}_{\rm {BG \it j}}} / {35}}
\label{2222}
\end{equation}
As for the Tibet observation, $\eta$ deviates from 1 at an order
of $10^{-3}$ depending on the Decl. of the on-source window.
 
Finally, the corrected estimation of the number of background events is obtained as

\begin{equation}
{N^{\ast}}_{\rm BG}= {\eta}N_{\rm {BG}} \label{3333} \end{equation}

It is clear that such a correction is free from the systematic uncertainty due to 
the slow variation of instrumental and environmental effects.
Denoting the number of excess events in an on-source window as $N_{\rm S}$, its uncertainty, $\Delta N_{\rm S}$, is given by ${\Delta N{\rm s}}^{2}={\Delta N_{\rm {ON}}}^{2}+{\Delta N^*_{\rm {BG}}}^{2}$, where ${\Delta N_{\rm {ON}}}^2$ is well approximated by $N_{\rm {BG}}^{*}$, and $\Delta N_{\rm {BG}}^{*}$ is
propagated from the statistic error of $\eta$ and $N_{\rm BG}$, according to
Equation (3), $\Delta {N_{\rm {BG}}^{*}}^2={N_{\rm {BG}}^{*}}^2(\frac{{\Delta\eta}^2}{\eta^2}+\frac{{\Delta N_{\rm {BG}}} ^2}{N_{\rm {BG}}^2})$.
Considering the number of off-source window and dummy on(off)-source window, it is not difficult to calculate $\Delta N_{\rm {BG}}^{*}$ as
${N_{\rm {BG}}^{*}}^2[(\frac{1}{35}+\frac{1}{350})(\frac{1}{N_{\rm {ON}}})+ (\frac{1}{10})(\frac{1}{N_{\rm {ON}}})]=0.1314{N_{\rm {BG}}^{*}}$.
Then we obtain
 $\Delta N_{\rm S}=1.0637\sqrt{N_{\rm {BG}}^{*}}$.
Because both
$N_{\rm {ON}}$ and $N^{*}_{\rm {BG}}$ are statistically large numbers, the significance value can  be simply calculated as
\begin{equation}
S = \frac{N_{\rm ON} - N^{\ast}_{\rm BG}}{1.0637\sqrt{N^{\ast}_{\rm BG}}} \label{4444} \end{equation}

\subsection{Method II (all distance equi-zenith angle method)}

Unlike the method described in the previous sub section, this method attempts to exploit the statistics as much as possible from two approaches. 

First, shapes of non-uniformly distributed azimuth angle distribution from whole data set are used to do the azimuth correction. To be more precise, after event selection, for each $1^{\circ}$ zenith angle interval between $0^{\circ}$ to $40^{\circ}$, azimuth angle distributions are filled and normalized, so that the averaged function value is one. Inversing the function and using it as an event weight would keep the total number of events unchanged but make the azimuth angle distribution flat.

Second, all events in the equi-zenith angle belt except those inside the on-source window are taken as off-source events. In this case the large-scale anisotropy effect can no longer be ignored, which leads to the attempt to fit simultaneously the relative CR intensity over the sky within the detector's field of view. 

The idea of this method is that at any moment, for all directions, if we scale down (or up) the number of observed events by dividing them with their relative CR intensity, then statistically, those scaled observed number of events in a zenith angle belt should be equal anywhere. A $\chi^2$ function can be built accordingly and the relative intensity of CRs in each  direction can be solved by minimizing the $\chi^2$ function.

As in the previous method, the celestial space from 0$^\circ$ to 360$^\circ$ in
R.A. and from 0$^\circ$ to 60$^\circ$ in Decl. are binned into cells with a bin size of 0.1$^\circ$ in both the R.A. and Decl. directions. In the observer's coordinates, the zenith angle $\theta$ is divided from 0$^\circ$ to 40$^\circ$ by a step size of 0.08$^\circ$, and the azimuth angle $\phi$ is binned by a zenith angle dependent bin width (0.08$^\circ$/ sin($\theta$)). For every local sidereal time (LST) interval bin (24s) $m$, a cell in ($\theta$, $\phi$) space $(n,l)$ is mapped to a celestial cell $(\alpha_i,\,\delta_j)$, i.e. R.A. bin $i$ and Decl. bin $j$, through two discrete coordinate transformation functions $\imath (m,n,l)$ and $\jmath (m,n,l)$. Therefore, at a certain LST bin $m$, the number of events
accumulated in zenith angle bin $n$ and azimuth angle bin $l$ is directly related to the CR intensity $I(i,j)$ in cell $(i,j)$ of celestial space.

Denoting the number of observed events after azimuth angle correction as $N_{\rm {OBS}}(m,n,l)$, and the relative intensity of CR as $I$(i,j), the equi-zenith angle condition leads to the following $\chi^2$ function:
\begin{equation}
\chi^2 \,=\, \sum_{m,\,n,\,l}\,
\frac{\;\Big[\,N_{\rm {OBS}}(m,\,n,\,l)\;\Big/\;I(i,\: j)\,-\,
{\sum_{\,l'\neq \,l}\Big( N_{\rm {OBS}}(m,\,n,\,l')\;/\;I(i',\: j')}\Big)\,\Big/\,{\sum_{\,l'\neq \,l} 1}\;\Big]^2}
{\; N_{\rm {OBS}}(m,\,n,\,l)\;\Big/\;{I^2(i,\: j)}\,+\, {\sum_{\,l'\neq \,l}\Big( N_{\rm {OBS}}(m,\,n,\,l')\;/\;I^2(i',\: j')}\Big)\,\Big/\,({\sum_{\,l'\neq \,l} 1})^2 }\,
\end{equation} \\
Here, $(i,j)$ is mapped from $(m,n,l)$ by the above-mentioned transformation functions.

From equation (5), $I(i,\: j)$ and its error $\Delta I(i,j)$ can be
solved numerically by the iteration method. As the dimensionless variable
$I(i,j)$ is only a relative quantity, and because the detecting
efficiency along the Decl. direction can not be absolutely calibrated, we apply
the constraint that averaged $I(i,j)$ in each Decl. slices is equal to one.

On the other hand, $N(i,j)$, the number of observed events in the celestial cell
$(i,j)$ is simply a summation of the relevant number of events
counted in local coordinates $N_{\rm {OBS}}(m,n,l)$, i.e., 
$N(i,j)= \sum\limits_{ \imath (m,n,l) = i \atop \jmath (m,n,l) = j } N_{\rm {OBS}}(m,n,l)$. 

As the relative intensity of the CR contains the contribution from large-scale anisotropy of the CR, to remove this effect and to isolate the contribution from point sources, the measured CR intensity in a $10^{\circ}$ belt along Decl. is projected to the R.A. direction, in $8^{\circ}$ bin size. The center of the belts moves from $0^{\circ}$ to
 $60^{\circ}$, in $1^{\circ}$ step size. By smoothing those curves and subtracting them from the corresponding belt of $1^{\circ}$ width, we obtain a CR intensity, $I_{\rm {CORR}}(i,j)$, that is corrected for the anisotropy effect.

With this corrected CR intensity,
the number of excess events and their uncertainties in cell $(i,j)$ can be
calculated as
\begin{equation}
N_{\rm {S}}(i,j)=\Big(I_{\rm {CORR}}(i,j)-1\Big)\cdot N(i,j)\,/\,I_{\rm {CORR}}(i,j)
\end{equation}
\begin{equation}
\triangle N_{\rm {S}}(i,j)=\Delta I_{\rm {CORR}}(i,j) \cdot N(i,j)\,/\,I_{\rm {CORR}}(i,j)
\end{equation}

Taking into account the array's angular resolution, events are summed up from a cone with an axis pointing to the source direction, and the half opening angle is set as $0.9^{\circ}$ (for ${\rm {E}} > 3$ TeV) or $0.4^{\circ}$ (for ${\rm {E}} > 10$ TeV). All celestial cells with their centers located inside the cone contribute to the number of events as well as its uncertainty. Finally, the significance for an on-source window centered at cell $(i,j)$ can be calculated as

\begin{equation}
{\it S}(i,j) \,=\, \frac{\sum\limits_{(i',j') \in {\rm {CONE}}}\Big((I_{\rm {CORR}}(i',j')-1)\cdot N(i',j')\,/\,I_{\rm {CORR}}(i',j')\Big)}
{\sqrt{\sum\limits_{(i',j') \in {\rm {CONE}}}\Big(\Delta I_{\rm {CORR}}(i',j') \cdot N(i',j')\,/\,I_{\rm {CORR}}(i',j')\Big)^2}}\,
\end{equation}


\section{Results}

The significance distributions from all directions are shown in Figure 2a and Figure 2b
for the two analysis methods, respectively. In the case of method II, the large-scale anisotropy
must be measured and subtracted, the systematic uncertainty to the significance value due to this subtraction procedure is estimated to be 0.2 $\sigma$ by changing the bin size and the smoothing parameters. 
Figure 3 shows the intensity map containing
the contribution from the large-scale anisotropy with a demonstration of its
subtraction.

  The excellent agreement on the negative side with a normal distribution (Figure 2a, Figure 2b) indicates that the systematic effects 
are well under control for both analyses. As for the positive side, a wider shoulder exists with
significance values  greater than 4.0 $\sigma$. The dominant contributions are
 due to the stable $\gamma$-ray source Crab and transient source Mrk~421
which was in an active state
from Feb. 2000 through Oct. 2001 \cite{Amenomori2003}. After removing their contributions, in such a manner that 
those cells that are less than 2$^\circ$ distance
from the two sources are excluded, the dot-dashed histograms in Figure 2a and Figure 2b which show the
significance distribution from the rest of the cells
agree much better with a normal distribution. For reference, the Table 1 lists the directions of local maximal significance points with significance values greater than 4.5 $\sigma$
from either of the two methods, and those prominent directions contribute to the remaining small shoulder in Figure 2a and Figure 2b.

Each significance value listed in Table 1 is calculated by equation (4) and (8) which is supposed to be used for a predefined on-source window. However, none of our prominent directions comes from a predefined one, as described in section 3.1, a large number of overlapping on-source windows have been tried. In comparison with finding a high significance value from one predefined direction (e.g., Crab or Mrk421), such a procedure makes the probability being larger by a factor of about the number of trials, and therefore making the excess of on-source event number less significant. It should be mentioned that it is very difficult to count the number of trials exactly. To have order of magnitude estimation on the lower boundary of the number of trials, we can divide the solid angle of surveyed sky by the solid angle of one on source window, and get a number of about 7000 which still under estimates the number of trials. As an example, for the first prominent point in Table.1, the significance value is 5.3 $\sigma$ from method I(4.8 $\sigma$ from method II), enlarge the probability by 7000 to account the number of trials effect end up with only 3.3(2.5) $\sigma$ in significance. To distinguish this significance from the previously calculated one, we call the former calculated one the pre-trials significance ($S_{pre-trials}$). This number of trials effect can also be understood from another point of view as the following: not including the Crab and Mrk421, both methods observe four directions which have $S_{pre-trials}$ greater than 4.5$\sigma$. From pure background MC simulation, the probability to observe four or more such directions is equivalent to a fluctuation of only 1.4$\sigma$. In conclusion, after taking into account the number of trials effect, the existence of the high significance point in Table 1 is well consistent with the background fluctuation.

As can be seen in Table 1, two established sources Crab and Mrk~421 are detected in an angular distance not larger than $0.5^{\circ}$ from their nominal positions,
in agree with the $0.4^{\circ}$ positional uncertainty estimated for
the both methods.
One of our previously reported \cite{Cui2003} hot point with pre-trials significance value of 4.0 $\sigma$ located at ($304.15^\circ,36.45^\circ$) is not listed in Table 1, because its pre-trials significance value is not improved after we further include more data samples into this analysis which increase the number of events by 60\%: $4.1$ $\sigma$ from Method I and $4.0$ $\sigma$ from Method II. The interesting thing about this point is that it is close to the Cygnus arm and was found to be close to one of Milargo hot points \cite{Walker2004},
though the large angular distance of $1.5^{\circ}$ indicates that the
coincidence is not in favour of a point source hypothesis. Conclusive
results will rely on further more observation. As a cross-check, Milagro hot point at ($79.9^\circ,26.8^\circ$) \cite{Atkins2004} is investigated in this work: the maximum pre-trials significant point is found at ($79.3^\circ,25.9^\circ$) with $1.3$ $\sigma$ from Method I, and ($79.9^\circ,26.3^\circ$) with $1.7$ $\sigma$ from Method II. In another word, no strong confirmation was obtained. On the other hand, except the agreement measurement
with comparable sensitivity on Crab and Mrk~421 from both observatories,
 no significant prominent direction from this work matches with
 Milagro hot points \cite{Atkins2004} within the allowed source positional uncertainties. When not considering the Crab and Mrk~421, we think the marginal coincidence
and refutation of other hot points between Tibet Air Shower and MILAGRO could be
attributed to the limited statistics and sensitivity of the two experiments. Whether these hot points are just due to statistical fluctuation or due to new TeV gamma sources will become clearer with the future improved statistic of observational data.

As described previously, the significances from all other directions other than Crab and Mrk~421 are not high enough to definitely claim any existence of new TeV $\gamma$-ray point source,
 we set a 90$\%$ confidence level (CL) upper flux limit for all
directions in the sky, except at the positions of Crab and Mrk~421. The upper limit on the number of events at the 90\% CL is calculated firstly
 according to the
number of excess events and its uncertainty for each cell following the statistic method given by 
Helene \cite{Helene1983}.
Then the effective detection area of Tibet Air Shower array is evaluated by 
full MC simulation assuming a Crab-like $\gamma$-ray
spectrum ${\rm {E}}^{-\beta}$ with $\beta = 2.6$ for a set of Decl. values
(0.0$^{\circ}$, 10.0$^{\circ}$, 20.0$^{\circ}$, 30.0$^{\circ}$, 40.0$^{\circ}$, 50.0$^{\circ}$, 60.0$^{\circ}$) and interpolated to other
Decl. values between $0^{\circ}$ and $60^{\circ}$.
Taking into account the live time from observational data taking, we derive the 90\%
 CL flux upper limit. Figure 4 shows the maps of flux upper limits at 90\% CL for energy ${\rm {E}} > 3$ TeV and for ${\rm {E}} > 10$ TeV (from Method I). Figure 5 shows the average flux upper limit along the right ascension direction as a function of Decl., which varies between
$1.5-3.2\times10^{-12} {\rm cm}^{-2}{\rm s}^{-1}$ for ${\rm {E}} > 3$ TeV
and
$0.2-0.4\times10^{-12} {\rm cm}^{-2}{\rm s}^{-1}$ for ${\rm {E}} > 10$ TeV.
Current integrated limits for energy above 10 TeV are the world-best ones 
which improve
 the results obtained from AIROBICC \cite{Aharonian2002a} for energy
above 15 TeV.
 For energy above 3 TeV, limits from this
work are comparable with those obtained by Milagro \cite{Atkins2004} for energy
above 1 TeV.
The same procedure is applied to the cases of other power law indices for energy above 3 TeV and above 10 TeV
. The corresponding average flux limit for
the other indices can be found in Figure 6 (from method II). Depending on the Decl. and power law index of the candidate source, the integral flux limits lie within $1.5-3.8\times10^{-12} {\rm cm}^{-2}{\rm s}^{-1}$ (${\rm {E}} > 3$ TeV) and $0.2-0.48\times10^{-12} {\rm cm}^{-2}{\rm s}^{-1}$  (${\rm {E}} > 10$ TeV).


\section{Conclusion}

A northern sky survey for the TeV $\gamma$-ray point sources in a Decl. band
between $0^{\circ}$ to $60^{\circ}$
was performed using data of the Tibet HD and the Tibet III air shower
array obtained from 1997 to 2001 with two independently developed analysis methods. The established Crab and
Mrk~421 were observed.
This indicates that the Tibet air shower array is a sensitive apparatus
for TeV $\gamma$-ray astronomy and has a potential to find
flare type $\gamma$-ray sources, such
as BL-Lac type AGNs. In addition, the 6 other prominent directions
were selected. However, more data will be needed before we can confirm or rule out these candidates.

With the exception of Crab and Mrk~421, 90\% CL flux upper limits are obtained from the rest of the positions
under the hypothesis that a candidate point source are in power law spectra, with indices varying from $2.0$ to $3.0$. The integral flux limits lie within $1.5-3.8\times10^{-12} {\rm cm}^{-2}{\rm s}^{-1}$ (${\rm {E}} > 3$ TeV) and $0.2-0.48\times10^{-12} {\rm cm}^{-2}{\rm s}^{-1}$  (${\rm {E}} > 10$ TeV) depending on the Decl. and power law index of the candidate source. 

\acknowledgments

This work is supported in part by Grants-in-Aid for Scientific Research
on Priority Areas (712) (MEXT) and by Scientific Research (JSPS) in Japan,
and by a grant for International Science Research from the Committee of the Natural
Science Foundation and the Chinese Academy of Sciences in China.

\clearpage

\begin{table}
\caption{Prominent directions with at least one pre-trials significance value greater than $4.5$ $\sigma$ from either of two methods. Each pre-trials significance values are calculated by the equation (4) or (8).
}
\begin{tabular}{c  c  c  c  c  c  c  c  c }
\hline
\hline
No.  &  Method & R.A.($^\circ$)  &  Decl.($^\circ$)  &  $N_{ON}$ &  $N_{OFF}$ & $N_S$ &  $\Delta N_s$  & $S_{pre-trials}$ \\
[0.5ex]
\hline
  & I & 69.95\tablenotemark{\star} \tablenotetext{\star}{ The last digital number in column of R.A. and Decl. are due to the way we divide the bin in the analyses. } & 12.05 & 828557.0 & 823428.5 & 5128.5 & 965.2 & 5.3    \\
\raisebox{1.5ex}[0pt]{1}  & II & 70.25 & 11.95 & 825589.0 & 821232.4 & 4356.6 & 906.2 & 4.8 \\
\hline 
 & I & 70.45 & 18.15 & 997441.0 & 992485.7 & 4955.3 & 1059.7 & 4.7   \\ 
\raisebox{1.5ex}[0pt]{2} & II & 70.45 & 18.05 & 994477.0 & 990343.3 &  4133.7 & 995.2 &  4.2  \\
\hline
 & I & 83.55 & 22.25 & 1087857.0 & 1082101.0 & 5756.0 & 1106.5 & 5.2  \\ 
\raisebox{1.5ex}[0pt]{3}  & II & 83.35 & 21.85 & 1079728.0 & 1074562.6 & 5165.4 & 1036.6 & 5.0   \\ 
\hline
& I & 88.85 & 30.25 & 1013683.0 & 1008916.9 & 4766.1 & 1068.4  & 4.5   \\ 
\raisebox{1.5ex}[0pt]{4} & II & 88.85 & 30.25 & 1185535.0 & 1179647.4 & 5887.6 & 1086.1 &  5.4 \\
\hline
  & I & 165.65 & 38.45 & 1170565.0 & 1164237.2 &  6327.8 & 1147.7  & 5.5  \\ 
\raisebox{1.5ex}[0pt]{5}  & II & 165.55 & 38.45 & 1170632.0 & 1164908.8  & 5723.2 & 1079.3 & 5.3  \\
\hline
 & I & 221.65 & 32.85 & 1062942.0 & 1057488.1 & 5453.9 & 1093.9  & 5.0   \\
\raisebox{1.5ex}[0pt]{6} & II & 221.75 & 32.75 & 1191295.0 & 1186765.1 & 4529.9 & 1089.4 &  4.2  \\
\hline
  & I & 286.65 & 5.45 & 608892.0 & 604934.2 & 3957.8 & 827.3  & 4.8    \\ 
\raisebox{1.5ex}[0pt]{7}  & II & 286.65 & 5.55 &  612041.0  & 608322.9 &  3718.1 &  780.0 &  4.8  \\
\hline
& I & 309.85 & 39.65 & 1146305.0 & 1141238.9 & 5066.1 & 1136.3  &  4.5  \\  
\raisebox{1.5ex}[0pt]{8} & II & 309.95 & 39.65 & 1146636.0 & 1141544.7 & 5091.3 & 1068.4 & 4.8 \\
\hline
 
\hline
\end{tabular}
\tablecomments{The meaning of the columns is (from left to right): sequence of prominent direction, analysis method, right ascension (J2000), declination (J2000), number of measured events in ON source window ($N_{ON}$), number of background events ($N_{OFF}$), event number excess in ON source window ($N_S=N_{ON}-N_{OFF}$), uncertainty on the event number excess ($\Delta N_s$), and pre-trials significance $S_{pre-trials}$ of deviation of $N_{ON}$ from $N_{OFF}$. }

\end{table}

\clearpage

\begin{figure}
  \includegraphics[height=16cm, width=16cm]{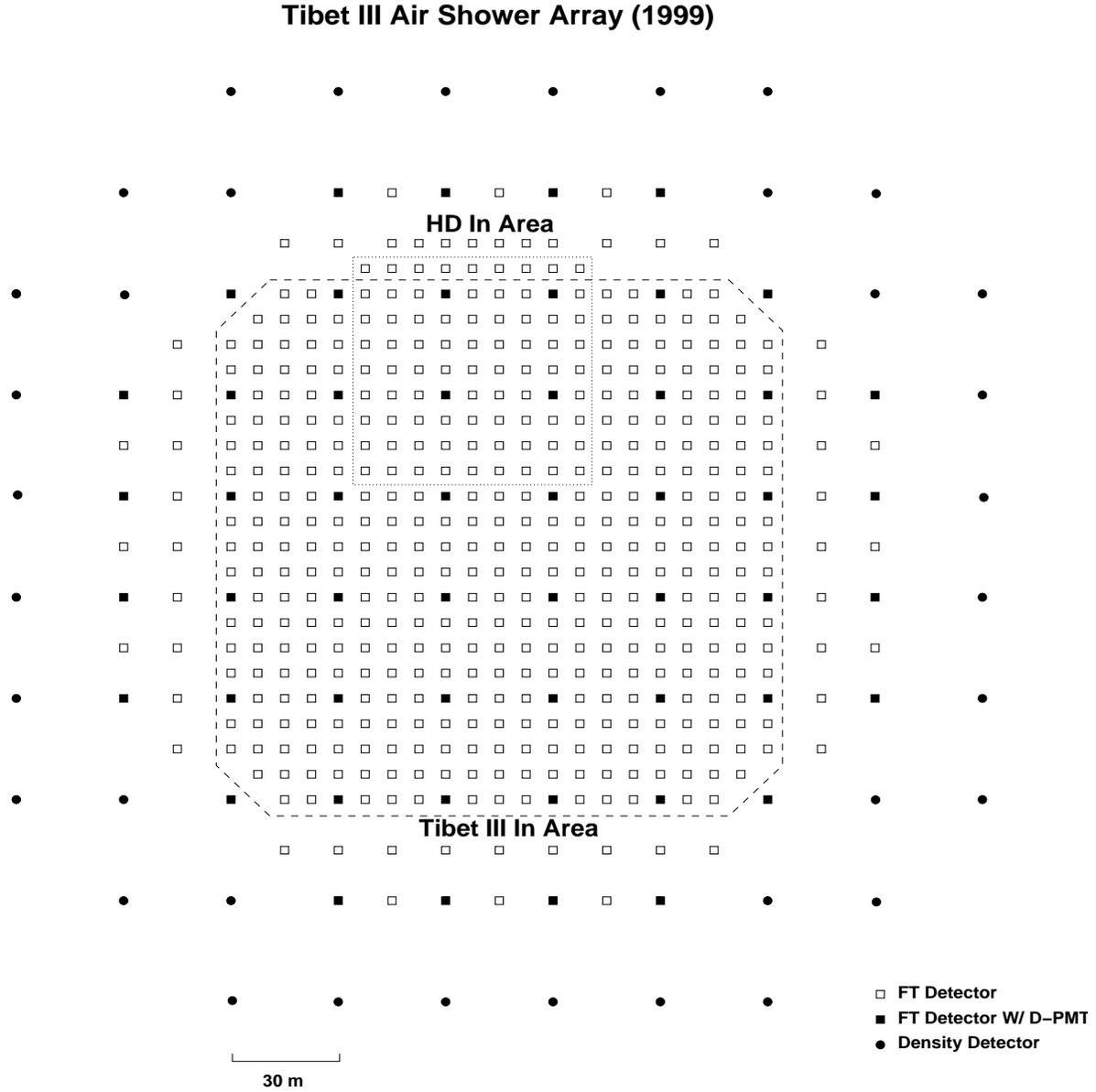}
  \caption{ Schematical view of the Tibet III array operating at Yangbajing. Open squares: FT-detectors; filled squares: FT-detectors with a wide dynamic range PMT; filled circles: density detectors with a wide dynamic range PMT. We select air shower events whose cores are located within the detector matrix enclosed with the dashed line.\label{TIBET_ARRAY}}
\end{figure}

\begin{figure}
  \includegraphics[height=8.5cm, width=8cm]{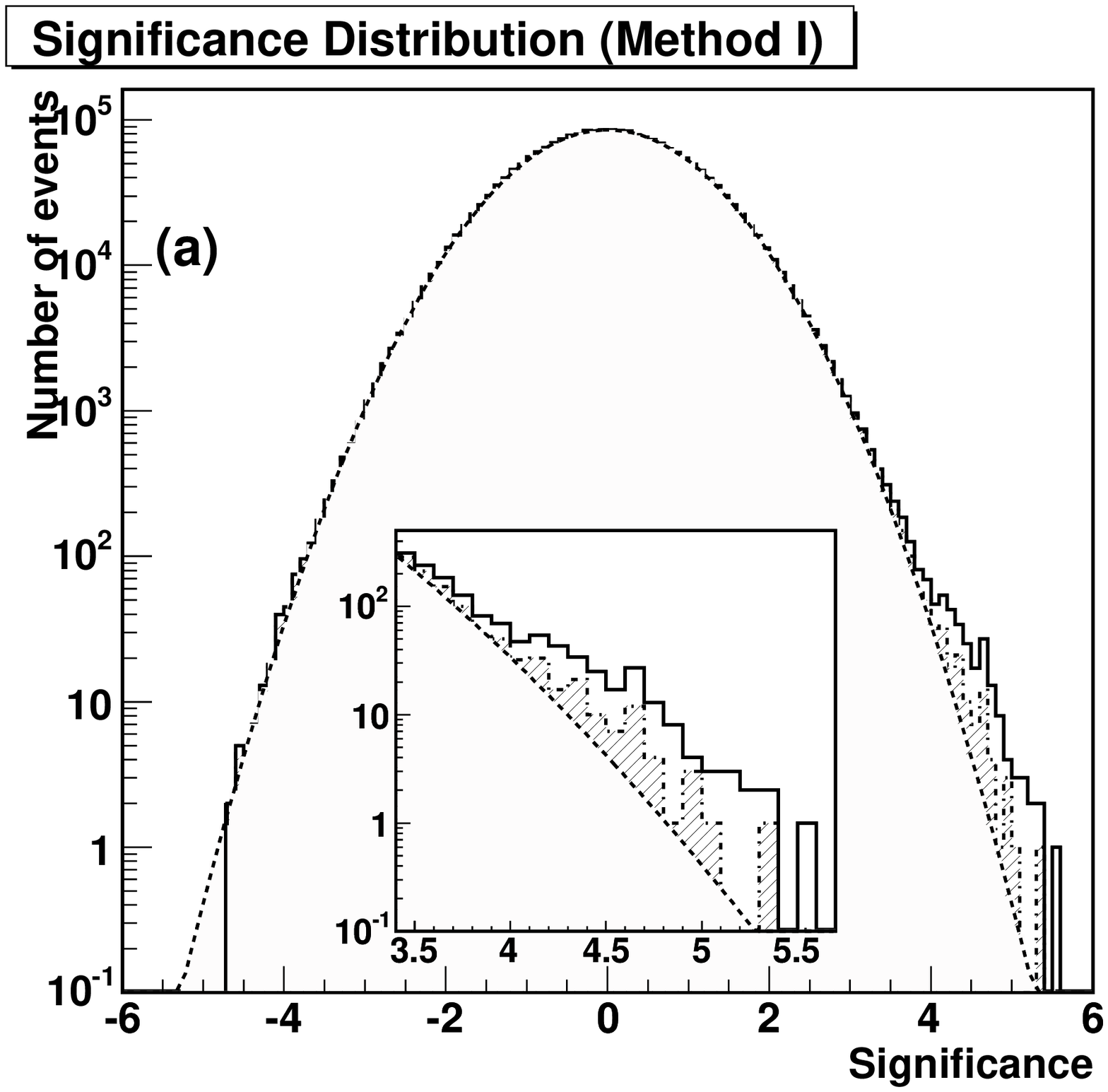}
  \includegraphics[height=8.5cm, width=8cm]{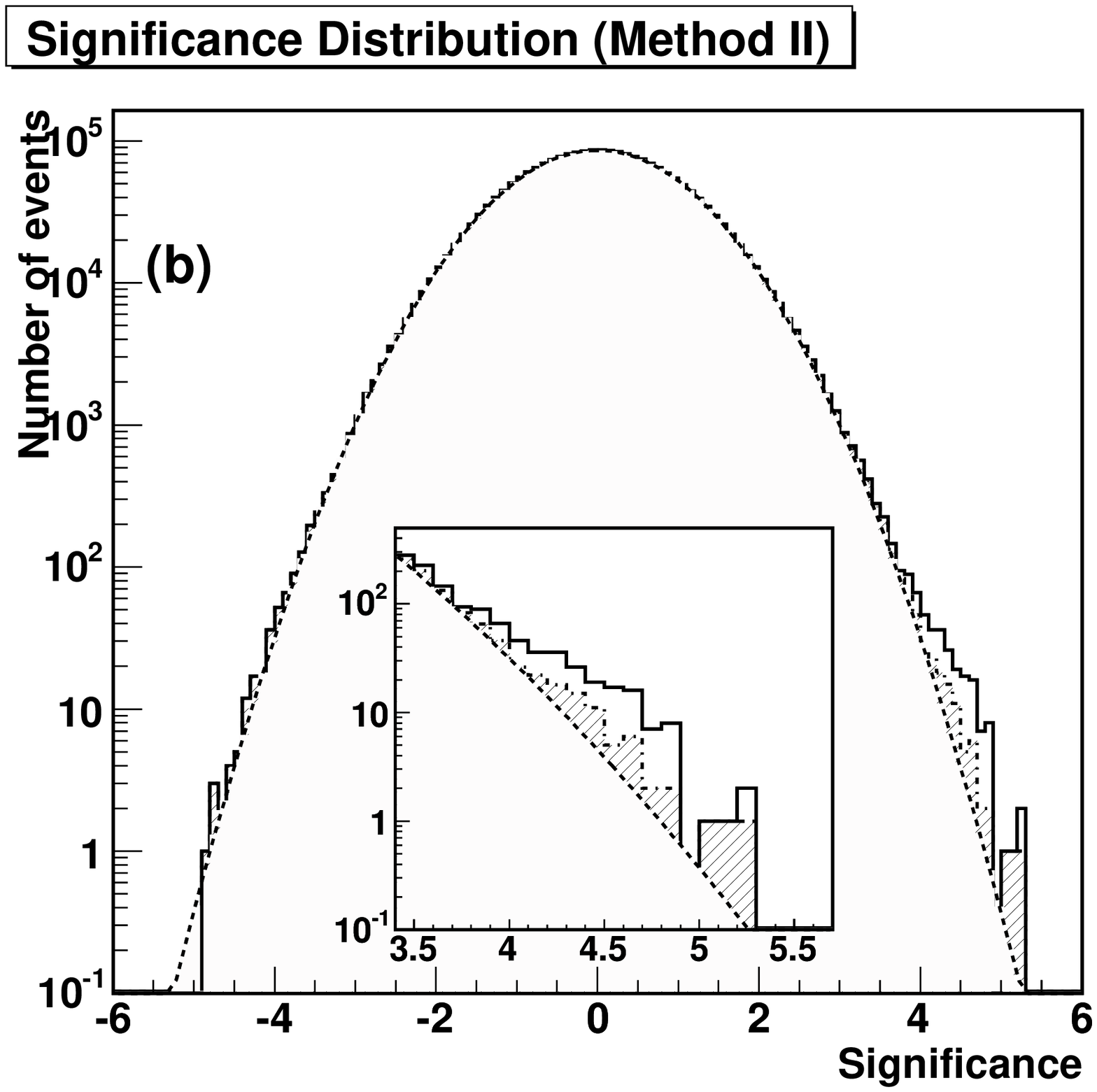}
  \caption{The pre-trials significance distribution of all directions on the sky map from (a) Method I and (b) Method II. It should be mentioned that not all directions are statistically independent as the bin size is smaller than the angular resolution. The solid line is derived from all cells defined in analyses. The dot-dashed histogram with shading excludes those cells which have a distance to Crab or Mrk~421 shorter than $2^\circ$. The dashed histogram represents the best Gaussian fit to the data. \label{sgnf_DIS}}
\end{figure}

\begin{figure}
  \includegraphics[height=10cm, width=16cm]{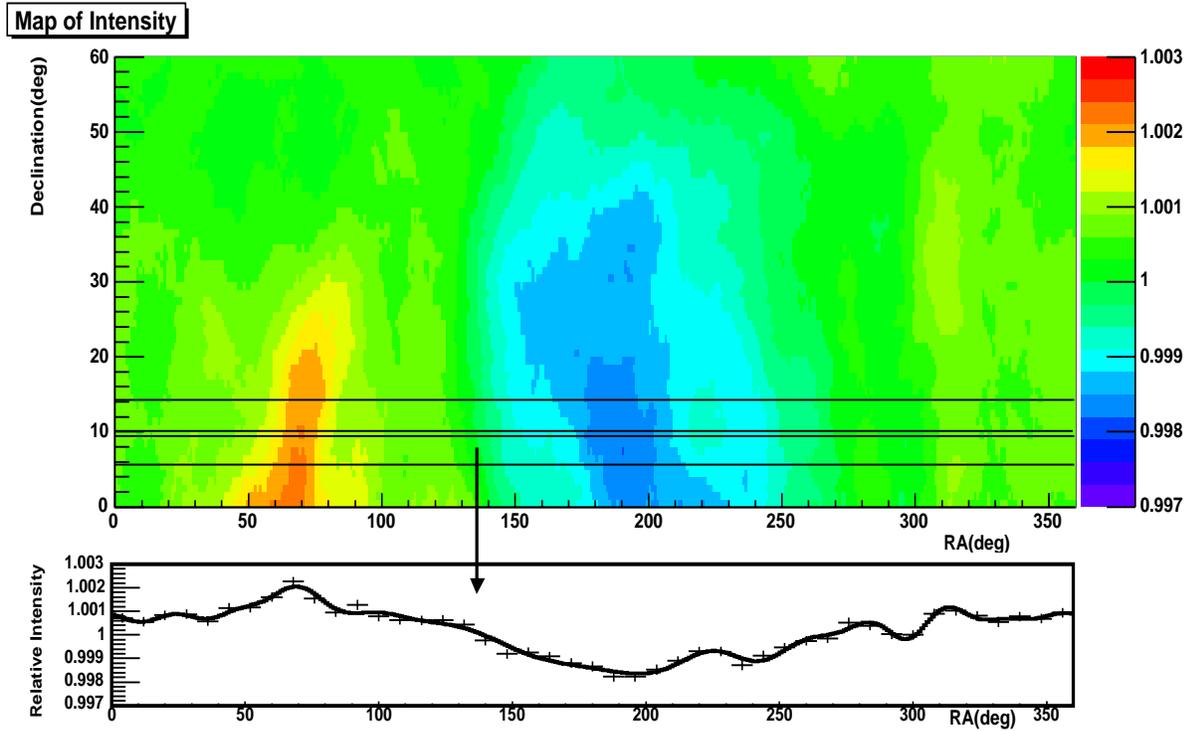}
  \caption{Upper panel: Map of the relative intensity of CR in the surveyed northern sky.
           Lower panel: A demonstration of the subtraction of the large-scale anisotropy of the CR. The anisotropy
                      distribution in the narrow belt centered at $9^{\circ}$ in Decl. is estimated by projecting the 
                      intensity measurement from a wider belt (a $10^{\circ}$ width is chosen) in the R.A. direction with 
                      a bin size of $8^{\circ}$. This distribution is further smoothed and then subtracted from the 
                      narrow belt.
  \label{intensity_2}}
\end{figure}

\clearpage

\begin{figure}
  \includegraphics[height=8cm, width=16cm]{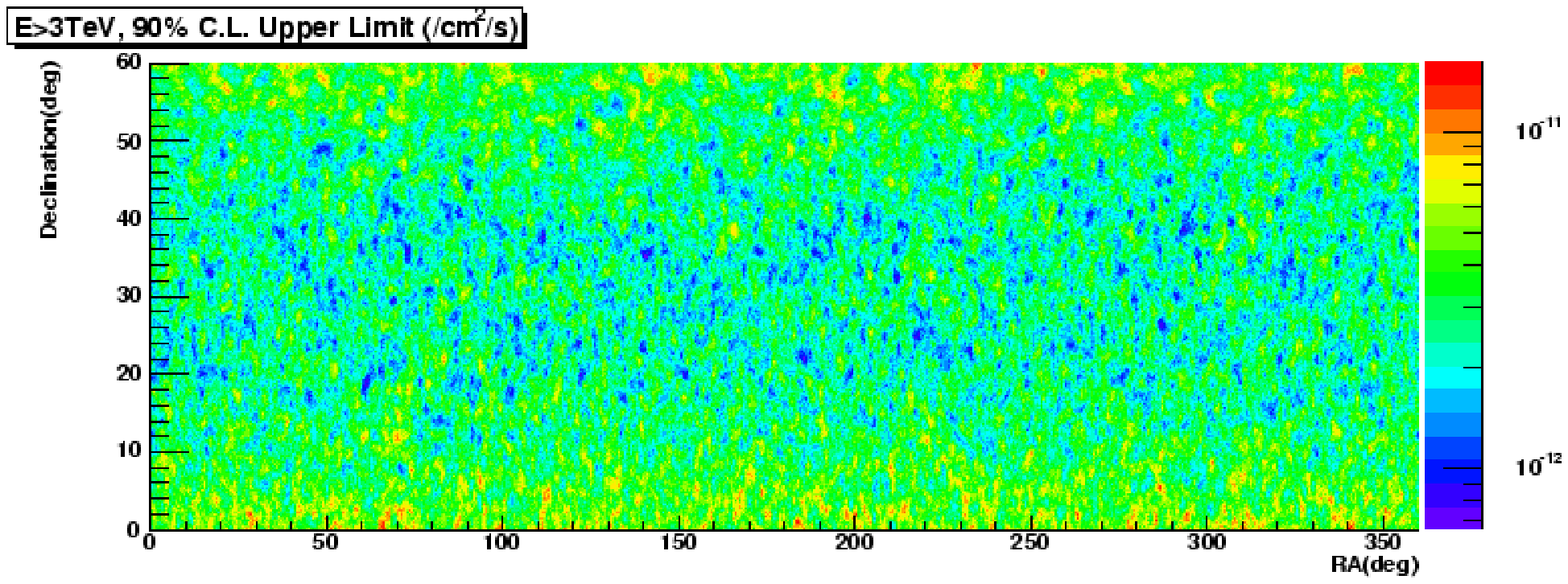}
  \includegraphics[height=8cm, width=16cm]{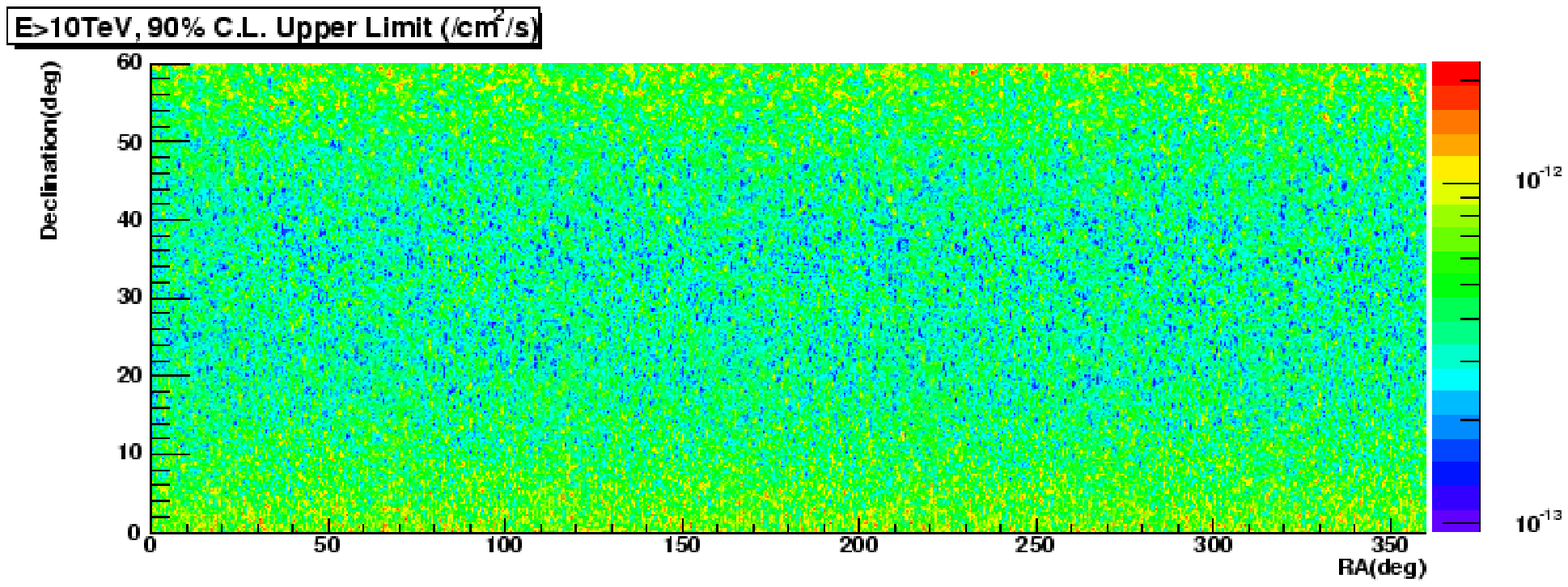}
 
\caption{
 Maps of 90\% CL flux upper limit assuming an energy spectrum of ${\rm {E}}^{-2.6}$(from Method I). Upper panel: above 3 TeV; lower panel: above 10 TeV.
}
\end{figure}

\begin{figure}
  \includegraphics[height=16cm, width=16cm]{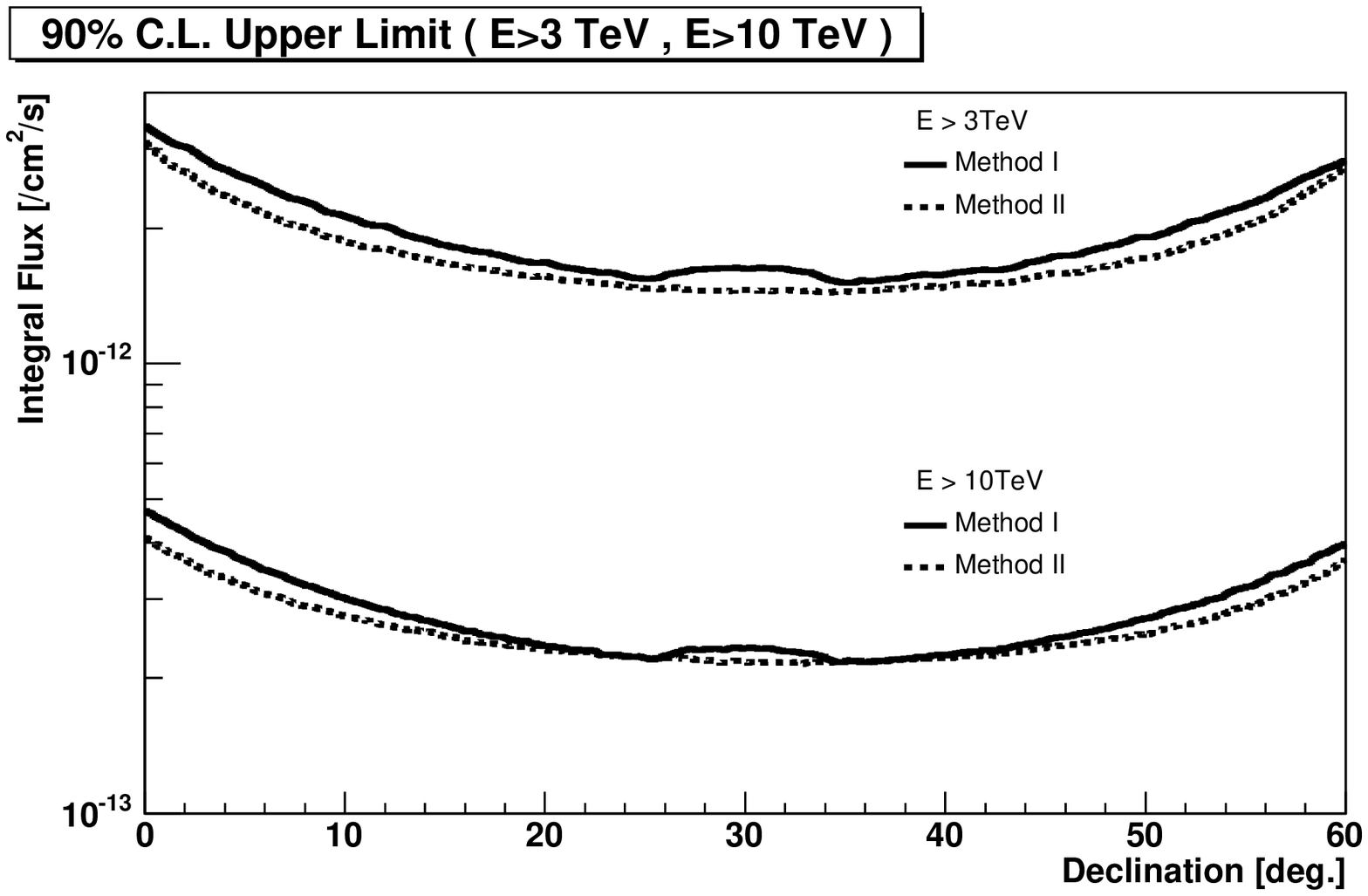}
 
\caption{
R.A. direction averaged 90\% CL upper limit on the integral flux above 3 TeV and 10 TeV for a Crab like point 
source, i.e., with an energy spectrum of ${\rm {E}}^{-2.6}$. The solid lines are obtained by Method I and the dashed lines by
Method II.
}
\end{figure}

\begin{figure}
  \includegraphics[height=16cm, width=16cm]{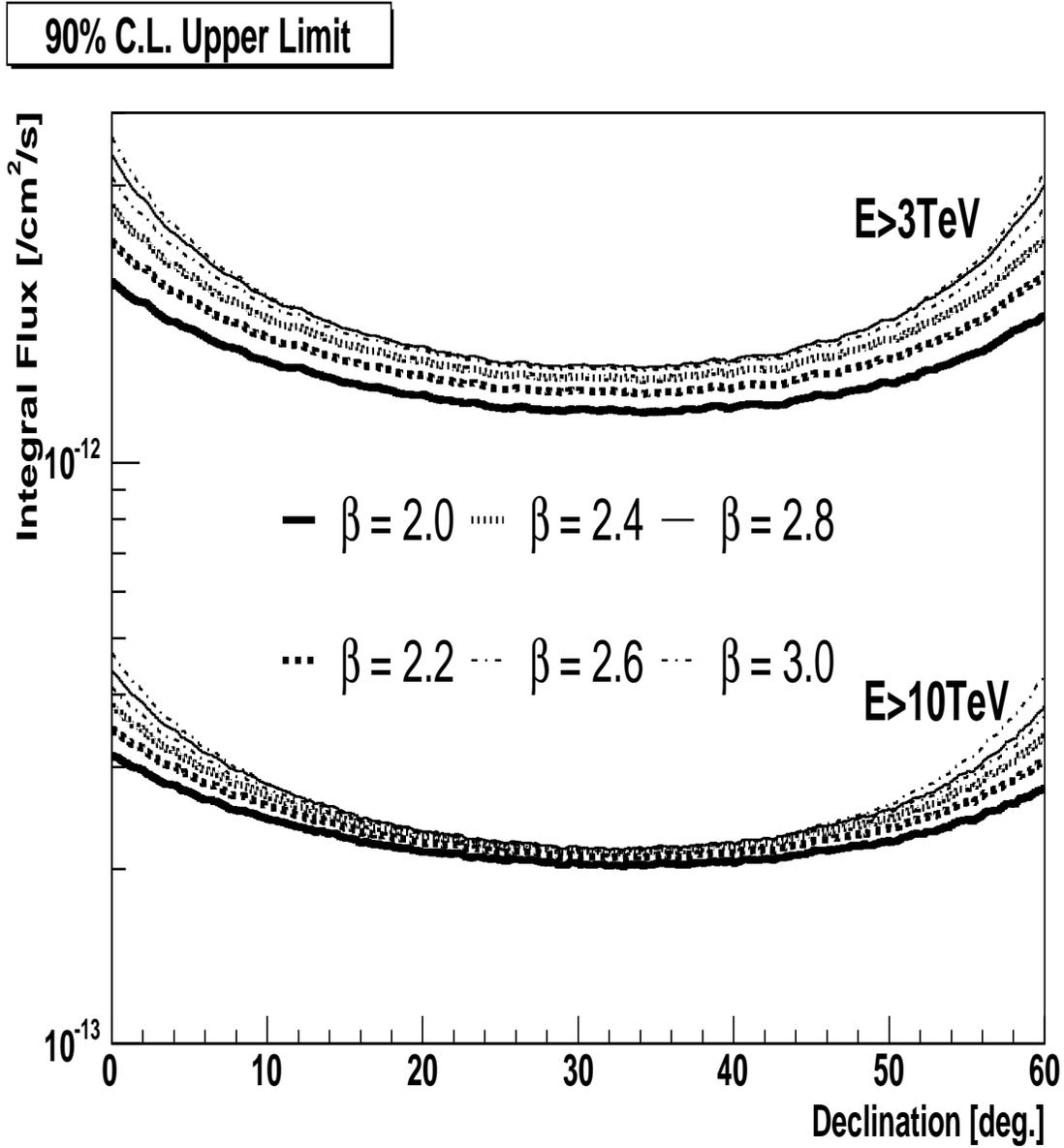}
 
\caption{
R.A. direction averaged 90\% CL upper limit on the integral flux above 3 TeV and 10 TeV for different indices of power law spectra. As an example, results from Method II are shown in this plot.
 \label{1sig_90CLUL_dec}
}
\end{figure}


\clearpage

\begin{thebibliography}{}
\bibitem[Alexandreas et al. 1991]{Alexandreas1991}
Alexandreas, D.E., et al., 1991, \apj, 383, L53-L56

\bibitem[Amenomori et al. 1990]{Amenomori1990}
Amenomori, M., et al. 1990, Nucl. Instrum. Methods Phys. Res., A288, 619

\bibitem[Amenomori et al. 1992]{Amenomori1992}
---------. 1992, \prl, 69, 2468

\bibitem[Amenomori et al. 1993]{Amenomori1993}
---------. 1993, Phy. Rev. D 47, 2675-2681

\bibitem[Amenomori et al. 1999]{Amenomori1999}
---------. 1999, \apjl, 525, L93

\bibitem[Amenomori et al. 2000]{Amenomori2000}
---------. 2000, \apj, 532, 302

\bibitem[Amenomori et al. 2001a]{Amenomori2001a}
---------. 2001a, Proc. 27th Int. Cosmic Ray Conf. (Hamburg), 6, 2544

\bibitem[Amenomori et al. 2001b]{Amenomori2001b}
---------. 2001b, AIP, CP558, High Energy Gamma-Ray Astronomy, P557

\bibitem[Amenomori et al. 2001c]{Amenomori2001c}
---------. 2001c,  Proc. 27th Int. Cosmic Ray Conf. (Hamburg), 2, 573

\bibitem[Amenomori et al. 2003]{Amenomori2003}
---------. 2003, \apj, 598, 242

\bibitem[Aharonian et al. 2002a]{Aharonian2002a}
Aharonian, F., et al., 2002a, \aap, 390, 39

\bibitem[Aharonian et al. 2002b]{Aharonian2002b}
---------. 2002b, \aap, 395, 803

\bibitem[Antoni et al. 2004]{Antoni2004}
Antoni, T., et al., 2004, \apj, 608, 865

\bibitem[Atkins et al. 1999]{Atkins1999}
Atkins, R., et al., 1999, \apjl, 525, L25-L28

\bibitem[Atkins et al. 2003]{Atkins2003}
---------. 2003, \apj, 595, 803-811

\bibitem[Atkins et al. 2004]{Atkins2004}
---------. 2004, \apj, 608, 680-685

\bibitem[Bloomer et al. 1988]{Bloomer1988}
Bloomer, S. D., Linsley, J.,\& Watson, A. A. 1988, J. Phys. G, 14, 645

\bibitem[Chen et al. 2004]{Chen2004}
Chen, X., et al., HEP \& NP, 2004, 28(10):1094-1098 (in Chinese)

\bibitem[Cui et al. 2003]{Cui2003}
Cui, S. W., et al., 2003, Proc. 28th Int. Cosmic Ray Conf. (Tsukuba), 4, 2315.

\bibitem[Falcone et al. 2003a]{Falcone2003a}
Falcone, A., et al., 2003a, Proc. 28th Int. Cosmic Ray Conf. (Tsukuba), 5, 2579

\bibitem[Hartman et al. 1999]{Hartman1999}
Hartman, R.C., et al., 1999, \apjs, 123, 79

\bibitem[Helene 1983]{Helene1983}
Helene, O., 1983, Nucl. Instrum. Methods Phys. Res., 212, 319

\bibitem[Horan et al. 2004]{Horan2004}
Horan, D., Weeks, T.C., 2004, New Astro. Rev. 48, 527-535

\bibitem[Hunter et al. 1997]{Hunter1997}
Hunter, S.D., et al., 1997, \apj, 481, 205

\bibitem[Ivanov et al. 1999]{Ivanov1999}
Ivanov, A.A., et al., 1999, JETP Lett., 69, 288 

\bibitem[Kasahara 2003]{Kasahara2003}
Kasahara, K., http://cosmos.n.kanagawa-u.ac.jp/EPICSHome/

\bibitem[Mckay et al. 1993]{Mckay1993}
Mckay, T.A., et al., 1993, \apj, 417:742-747

\bibitem[Sinnis et al. 2002]{Sinnis2002}
Sinnis, C. 2002, proceedings of American physical society and High energy Astrophysics of the AAS Mtg

\bibitem[Ong 2003]{Ong2003}
Ong, Rene A., 2003, The Universe Viewed in Gamma-Rays, 587 (Universal Academy Press, Tokyo).

\bibitem[V${\rm \ddot{o}}$lk 2003]{Voelk2003}
V${\rm \ddot{o}}$lk, H.J., 2003, Frontiers of Cosmic Ray Science, 29 ( Universal Academy Press, Tokyo )

\bibitem[Walker et al. 2003b]{Walker2003b}
Walker, G., et al., 2003b, Proc. 28th Int. Cosmic Ray Conf. (Tsukuba), 5, 2563

\bibitem[Walker et al. 2004]{Walker2004}
Walker, G., Atkins, R., Kieda, D., 2004, \apjl, 614, L93-L96

\bibitem[Wang et al. 2001]{Wang2001}
Wang, K. et al., 2001, \apj, 558, 477

\bibitem[Weeks et al. 1989]{Weeks1989}
Weeks, T.C., et al., 1989, \apj, 342, 379

\bibitem[Weeks 2003]{Weeks2003}
Weeks, T.C., 2003, Frontiers of Cosmic Ray Science, 3 ( Universal Academy Press, Tokyo )

\end{thebibliography}
\end{document}